\documentclass[12pt]{iopart}
\include{graphicx}

\begin{document}


To appear in The Astronomical Journal

\title[Observations at 7-mm of the Quadrupolar HH~111/121 Outflow]
{High Angular Resolution Observations at 7-mm \\
of the Core of the Quadrupolar HH~111/121 Outflow}


\author{Luis F. Rodr\'\i guez}
\address{Centro de Radioastronom\'\i a y Astrof\'\i sica, UNAM, Morelia 58089, M\'exico}
\ead{l.rodriguez@astrosmo.unam.mx}

\author{Jos\'e M. Torrelles}
\address{Instituto de Ciencias del Espacio (CSIC)-IEEC, Facultat de F\'\i sica, 
Universitat de Barcelona, E-08028 Barcelona, Spain}
\ead{torrelles@ieec.fcr.es}

\author{Guillem Anglada}
\address{Instituto de Astrof\'\i sica de Andaluc\'\i a (CSIC), Apartado 3004, E-18080 Granada, Spain}
\ead{guillem@iaa.es}

%\and

\author{Bo Reipurth}
\address{Institute for Astronomy, University of Hawaii, Hilo, HI, USA}
\ead{reipurth@ifa.hawaii.edu}


%\altaffiltext{1}{Harvard-Smithsonian Center for Astrophysics,
%and 
%Submillimeter Array, 60 Garden Street, Cambridge, MA 02138, USA}


\begin{abstract}
We present sensitive, high angular resolution ($0\rlap.{''}05$) VLA 
continuum observations 
made at 7 mm of
the core of the HH~111/121 quadrupolar outflow.
We estimate that at this wavelength the continuum emission is dominated by dust, although
a significant free-free contribution ($\sim$30\%) is still present.
The observed structure is formed by two overlapping, elongated sources
approximately perpendicular to each other as viewed from Earth. 
We interpret this structure
as either tracing two circumstellar disks that exist around each of the protostars 
of the close binary source at the core of this quadrupolar outflow
or a disk and a jet perpendicular to it. Both interpretations have advantages and disadvantages,
and future high angular resolution spectroscopic millimeter observations
are required to favor one of them in a more conclusive way.

\end{abstract}

\section{Introduction}

It is generally accepted that most stars form in binary or multiple systems
(Lada \& Lada 2003).
Furthermore, in the case of low and intermediate-mass stars it is also
known that the process occurs with the presence of an accretion disk and
a collimated outflow. From these two facts it follows that binary disk-jet systems
should be common in regions of low-mass star
formation. However, in practice
these systems are hard to detect and identify and only a few 
have been studied in detail
(e. g. Rodr\'\i guez et al. 1998; Anglada et al. 2004; Monin et al. 2007).
Furthermore, it is still unclear if the members of a binary system will
both be able to maintain the disks and outflows that characterize
the formation of
single stars. For example, the binary stars that form L1551~IRS5 are
both believed to possess disks and jets (Rodr\'\i guez et al. 2003;
Lim \& Takakuwa 2006), while it is known that only one
of the stars that forms the SVS 13 close binary system is
associated with detectable circumstellar dust emission
that is probably tracing a disk (Anglada et al. 2004).

One of the most interesting cases of a binary source that is known to
exhibit independent outflows, most probably associated with
each of the components of the binary, is HH~111 (Reipurth et al. 1999).
Located in Orion at a distance of 414 pc (Menten et al. 2007), the optical HH 111 jet was
discovered by Reipurth (1989).
It is an extremely well-collimated jet, aligned approximately in the east-west
direction (at a PA of $\sim 97^\circ$) and whose knots move in the plane of the sky
with velocities
of the order of several hundred km s$^{-1}$ (Reipurth, Raga \& Heathcote 1992).
Reipurth, Bally \& Devine (1997) found that this optical jet is part of
a giant HH complex extending over 7.7 pc.
This giant HH complex is very straight, suggesting great stability over the
$10^4$ years of its lifetime. 

Near-infrared observations (Gredel \& Reipurth 1993; 1994)
revealed a second bipolar flow, named HH 121, that emerges from about the same
position as the optical outflow, and is aligned approximately in the
north-south direction (at a PA of $\sim 35^\circ$). This result 
suggested the presence of a close binary source
in this region.
Both the optical and the infrared outflows are detected as bipolar molecular outflows
(Cernicharo \& Reipurth 1996; Nagar et al. 1997; Lefloch et al. 2007).

At the center of the
quadrupolar outflow is the source IRAS 05491+0247 = VLA 1, a suspected class I binary
with a total luminosity of about 25 L$_\odot$ 
(e.g. Stapelfeldt \& Scoville 1993, Yang et al. 1997).
To advance in our understanding of this source a high angular resolution image
at millimeter wavelengths was needed to compare with the information available 
for the quadrupolar outflow.

\section{Observations}

%% In a manner similar to \objectname authors can provide links to dataset
%% hosted at participating data centers via the \dataset{} command.  The
%% second curly bracket argument is printed in the text while the first
%% parentheses argument serves as the valid data set identifier.  Large
%% lists of data set are best provided in a table (see Table 3 for an example).
%% Valid data set identifiers should be obtained from the data center that
%% is currently hosting the data.
%%
%% Note that AASTeX interprets everything between the curly braces in the 
%% macro as regular text, so any special characters, e.g. "#" or "_," must be 
%% preceded by a backslash. Otherwise, you will get a LaTeX error when you 
%% compile your manuscript.  Special characters do not 
%% need to be escaped in the optional, square-bracket argument.

The 7 mm continuum observations were made in the A configuration
of the VLA of the NRAO\footnote{The National Radio
Astronomy Observatory is operated by Associated Universities
Inc. under cooperative agreement with the National Science Foundation.}, during 
2006 February 10 and 15. The central frequency observed was
43.34 GHz and we integrated on-source for a total of
approximately 5 hours. The absolute amplitude
calibrator was 1331+305 (with an adopted flux density of 1.46 Jy)
and the phase calibrator was 0552+032 (with a bootstrapped flux density
of 0.84$\pm$0.01 Jy and 0.89$\pm$0.04 Jy, for the first and
second epoch, respectively). The phase noise rms was about 20$^\circ$
and 30$^\circ$, for the first and second epoch, respectively,
indicating good weather conditions. The phase center of the observations was at
$\alpha(2000) = 05^h~51^m~46\rlap.^s25;~\delta(2000) = +02^\circ~48{'}~29\rlap.^{''}6$.

The data were acquired and reduced using the recommended VLA procedures
for high frequency data, including the fast-switching mode with a
cycle of 120 seconds. The effective bandwidth of the observations
was 100 MHz.

%% In this section, we use  the \subsection command to set off
%% a subsection.  \footnote is used to insert a footnote to the text.


%% This section also includes several of the displayed math environments
%% mentioned in the Author Guide.

\section{Analysis}

\subsection{7 mm Data}

In Figure 1 we show the 7 mm image of the core of the HH 111/121 quadrupolar
outflow. An unusual structure is evident in the image. The structure can
be described as two overlapping, elongated sources
aligned in the plane of the sky and approximately perpendicular to each other. 
We have fitted the emission with two Gaussian ellipsoids using the
task JMFIT of the AIPS package (see Table 1).  
The data image and the model image are shown in Figure 2.
In what follows, we will interpret this 7 mm structure, as well as
that seen at 3.6 cm, in terms of combinations of ionized jet/dusty disk
scenarios. However, other possibilities such as the presence of
non-thermal components should be recognized.

For the purpose of comparison, we also show in Figure 1 the 3.6 cm
continuum image
obtained from the VLA data presented by Reipurth et al. (1999). 
These data have been recalibrated and the position of their phase calibrators
updated by displacements of order
$0\rlap.{''}16$ to the most recent values in the VLA calibrator catalog
to allow accurate astrometric comparison with the 7 mm image.
As first discussed by these authors, the 3.6 cm VLA image is
suggestive of a quadrupolar jet, with a common origin within $0\rlap.{''}1$ 
($\sim$40 AU).
The main jet is aligned approximately in the east-west direction
(that with a PA of $97^\circ \pm 1^\circ$ is taken to be the exciting agent of the extended 
optical HH 111 flow),
while the second jet is aligned approximately in the north-south direction
(that with a PA of $4^\circ \pm 4^\circ$ 
is taken to be the exciting agent of HH 121). The error in the position angle
of the north-south jet was obtained from our reanalysis
of the Reipurth et al. (1999) data. The quadrupolar structure of the
3.6 cm emission is more evident in the maximum entropy reconstruction of
the image shown by Reipurth et al. (1999). From this image we crudely estimate that 
about 70\% (0.7$\pm$0.2 mJy) of the total 3.6 cm emission (1.0 mJy; see Table 2) 
comes from the east-west jet, while the
remaining 30\% (0.3$\pm$0.2 mJy) comes from the weaker north-south jet.

We are then faced with a question: since the position 
angles of the two nearly perpendicular structures observed at
7 mm are similar to those observed at 3.6 cm, are we seeing at
7 mm two jets, two disks, or a jet and a disk?
To address these interpretations, we first discuss additional
continuum observations of the region.
 
%There are two lines of argument that suggest that at 7 mm
%we are seeing dust emission from protoplanetary disks. The first
%is that at 3.6 cm the structure with the larger flux density is the
%east-west one, while at 7 mm this is true for the north-south structure
%(see Fig. 1 and Table 1). If we assume that the flux density of disks and
%jets is correlated, this favors matching the east-west structure at
%3.6 cm (the jet) with the north-south structure at 7 mm.

%A stronger argument comes from discussing the centimeter and millimeter
%continuum spectrum of the source, that we do in the following section.

\subsection{Flux Densities at Other Wavelengths} 

We searched in the literature and in the VLA archives for additional
data points of the continuum flux density of this source.
In Table 2 we present a summary of the total flux densities obtained for
the region. The observations are not taken simultaneously and
if time variations are present, this will lead to an erroneous
determination of the spectral indices.
We also note that the observations are taken at different angular
resolutions and that this may affect the comparison.
The VLA is sensitive only to emission in angular scales
smaller or comparable to $\sim$25 times its angular resolution, while for OVRO
the largest detectable angular scale is about 10 times the angular resolution.
In particular, the 7 mm data
was taken with the highest angular resolution (and thus the smallest
detectable angular scale) and 
we may be underestimating the total flux density at this
wavelength in comparison with the other observations, that
are all taken at lower angular resolutions. 
Lower angular resolution VLA observations at 7 mm are needed to 
determine the complete flux density of HH~111.
%The VLA archive data at 4.86 GHz was taken in 1994 September 21 
%in the CnB configuration. 

The total continuum flux density
at 7 mm is about 40\% larger than the addition of the
individual flux densities given in Table 1.
This difference results from the presence of faint, extended
7 mm emission that is not accounted for in the individual
Gaussian fits. Analysis of the residual (with the
Gaussian fits removed) image suggests that this faint emission seems 
to extend over a diameter
of $\sim0\rlap.{''}3$, but its nature is unclear.
As can be seen from Table 2 and in Figure 3,
the flux density is relatively flat in the centimeter range and rises rapidly above
$\sim$30 GHz. We have fitted the data points with the sum of two power laws
of the form:

$$S_\nu(total) = S_{8.4~GHz}(free-free) (\nu/8.4~GHz)^{0.3} +  
S_{43.3~GHz}(dust) (\nu/43.3~GHz)^{2.5},$$

\noindent where the first term is intended to represent the contribution
of free-free emission from the ionized outflow and the second term
is intended to represent the contribution
of dust emission from the disks. 
Since we have four parameters to fit and only five data points,
this fit is only intended to show that the data can be reasonably
fitted with the sum of two power laws.
In Figure 3 we show the 
fit for $S_{8.4~GHz}(free-free)$ = 0.93 mJy and
$S_{43.3~GHz}(dust)$ = 3.6 mJy.
Under this interpretation, the 8.44 GHz emission is
clearly dominated by free-free,
with only $\sim$6\% of dust contribution. On the other hand,
the 43.34 GHz emission is dominated by dust,
with a $\sim$30\% contribution from free-free. 
%Even when the free-free 
%contribution is significant, we will interpret the 43.34 GHz emission
%as dominated by dust.
Following Rodr\'\i guez, Zapata, \& Ho (2007) and
Hunter et al. (2006), we estimate the total mass in 
the dust to be in the order of 0.1 $M_\odot$.
For this estimate a dust temperature of 45 K was used, following
the assumptions of Stapelfeldt \& Scoville (1993).

\subsection{Two jets?}

The excess of 7-mm emission with respect to the value expected from the
extrapolation of the
centimeter measurements rules out the possibility that at 7-mm we are
simply observing
two jets since we have estimated that a significant fraction ($\sim$70\%) of the 
7-mm emission is due to dust. However, we cannot fully rule out the possibility that
we are seeing a core dominated by dust emission with the extended
emission dominated by free-free emission. 
%Given the limitations
%of the data, we will assume that the dust and free-free emissions present
%have a similar spatial distribution. 

\subsection{Two disks?}

This possible interpretation is suggested by the fact that at
3.6 cm the structure with the larger flux density is the
east-west one, while at 7 mm this is true for the north-south structure
(see Fig. 1 and Table 1). 
%At 3.6 cm the east-west structure has a flux density of
%0.8 mJy, while at 7-mm the north-south structure has a flux density
%of 2.2 mJy.  
Rodr\'\i guez et al. (2008), from a comparison between the momentum
rate in the molecular outflow and the radio continuum emission from the jet,
have argued that high-mass young stars can be understood as a scaled-up
version of low-mass young stars. Assuming that this conclusion is valid also for the
parameters of jets and disks, and we tentatively assume 
that the flux density of disks and
jets is correlated, this favors matching the east-west structure at
3.6 cm (the jet) with the north-south structure at 7 mm (the disk),
and the north-south structure at
3.6 cm (the second jet) with the east-west structure at 7 mm
(the second disk), forming a binary disk-jet system
as that observed in L1551~IRS5 (Rodr\'\i guez et al. 2003).
 
This interpretation faces, however, three difficulties.
The first difficulty is that the axes of the proposed disk-jet structures
are not, as expected, nearly perpendicular.
The east-west structure at
3.6 cm has a PA of $97^\circ \pm 1^\circ$, while the
north-south structure at 7 mm has a PA of $23^\circ \pm 6^\circ$,
having a relative angle of $74^\circ \pm 6^\circ$ between them.
Furthermore, the north-south structure at
3.6 cm has a PA of $4^\circ \pm 4^\circ$, while the
east-west structure at 7 mm has a PA of $116^\circ \pm 7^\circ$,
having a relative angle of $112^\circ \pm 8^\circ$ between them.

The second difficulty is related to the 
stability of the proposed binary disk system.
From Table 1, we estimate that the centroids of the two 7 mm structures are
separated by $0\rlap.{''}036 \pm 0\rlap.{''}014$ or $\sim$15$\pm$6 AU at a 
distance of 414 pc.
We expect the disks in a 
binary system to be truncated at radii of order 
a fraction of the binary separation (Armitage, Clarke, \& Tout 1999; Papaloizou \& Pringle 1987;
Pichardo, Sparke, \& Aguilar 2005).
We then expect the radii of the disks to be compact,
$\leq$15 AU (assuming that the observed separation is
comparable to the true separation).
In contrast, the observed radii (see Table 1) are
significantly larger,
in the order of 35 AU. 

The final difficulty is that, if we assume masses of order 1 $M_\odot$
for the stars in the binary system, an orbital period of order
40 years is expected for a separation of $\sim$15 AU. 
The expected timescale in which strongly misaligned binary
disks should be brought into rough alignment by tidal torques is
about 20 orbital periods (Bate et al. 2000), which is only
800 years if the 15 AU separation is real.
With this relatively short timescale it is difficult to
explain the lifetime of 10$^4$ years of the HH 111 jet
and its remarkable stability over this period.
On the other hand, it should be noted that Reipurth et al.
(1992) have suggested that the separation of bright knots in the 
optical flow are consistent with a timescale of variation
of the central source of 40 years.

These last two difficulties are mitigated if we consider that the real separation is much larger
than the projected one. A physical separation of 105 AU would be consistent
with the values of 35 AU for tidally truncated radii. The orbital
period would now be about 800 years and the timescale for alignment
$\sim 1.6 \times 10^4$ years, consistent with the lifetime of
the system. 
However, if we define the physical separation as $r$, the
probability of observing it at a projected separation 
of $r'$ or smaller is:

$$P(\leq r') = 1 - [1 - (r'/r)^2)]^{1/2},$$

\noindent that for $r'$ = 15 AU and $r$ =105 AU,
gives $P(\leq r') \simeq 0.01$.

We conclude that either we are observing the binary at an unlikely
orientation, that disks in binary systems are more stable
and independent than previously thought, or that the interpretation is
incorrect.

\subsection{One jet and one disk?}

This final interpretation implies that at 7-mm we are seeing a 
north-south structure dominated by dust emission that traces a
disk and an east-west structure that could be related with the jet
that powers the optical HH~111 outflow. A major advantage with this
interpretation is that the structures have a relative angle of
$93^\circ \pm 9^\circ$, very close to perpendicularity.
The east-west structure has a flux density of 0.7$\pm$0.2 mJy at 3.6 cm and
of 1.4$\pm$0.3 mJy at 7-mm. This results in a spectral index of 0.4$\pm$0.2, that is
consistent with the value expected for a thermal jet (e. g. Anglada 1996;
Eisl\"offel et al. 2000). 

However, this interpretation also presents some difficulties.
The east-west structure at 3.6 cm has a position angle
of $97^\circ \pm 1^\circ$, while the east-west structure at 7 mm has a position angle
of $116^\circ \pm 7^\circ$. In other words, the two structures are not nearly parallel, 
as expected if the 3.6 cm emission were the larger scale counterpart of the
7 mm jet emission.   
Under this interpretation, all the 7-mm emission is related to the
HH~111 disk-outflow system and there is no evidence at this wavelength of the
HH~121 system.
Of course, it should be pointed out that the dust and free-free emission
from the HH~121 system are probably present in the image at a low level
but that we cannot disentangle their presence given the modest signal-to-noise
ratio of the data.
% and could 
%help alleviate the lack of alignment between the structures. 

\section{Conclusions}

Our main conclusions follow:

1) Our high angular resolution ($0\rlap.{''}05$) VLA 
continuum 7 mm observations of  
the core of the HH~111/121 quadrupolar outflow
reveal a structure that can be described as two overlapping, elongated sources
that appear approximately perpendicular to each other
in the plane of the sky. 

2) We discuss possible interpretations for this structure and conclude that
the most viable ones are that we are observing two orthogonal disks
around separate protostars or 
a disk with a perpendicular jet. Both intepretations
have advantages and disadvantages,
and high angular resolution spectroscopic millimeter observations
(possible only in the future with the Atacama Large Millimeter Array)
are required to disentangle what is going on at the core of this quadrupolar
outflow.

\ack

We thank an anonymous referee for valuable suggestions. 
LFR acknowledges the support
of CONACyT, M\'exico and DGAPA, UNAM.
JMT and GA are supported by the MEC AYA2005-05823-C03 grant
(co-funded with FEDER funds). GA also acknowledges support from Junta de
Andaluc\'{\i}a.

\section*{References}
%\begin{thebibliography}{}
\begin{harvard}

\item[] Anglada, G. 1996, in ASP Conf. Ser. 93, Radio 
Emission from the Stars and the Sun, ed. A. R. Taylor \& J. M. Paredes 
(San Francisco: ASP), 3 

\item[] Anglada, G., Rodr\'\i guez, L. F., Osorio, M.,
Torrelles, Jos\'e M., Estalella, R., Beltr\'an, M. T., \&
Ho, P. T. P. 2004,
ApJ, 605, L137

\item[] Armitage, P. J., Clarke, C. J., \& Tout, C. A. 1999, MNRAS, 304, 425

\item[] Bate, M. R., Bonnell, I. A., Clarke, C. J., Lubow, S. H., Ogilvie, G. I., 
Pringle, J. E., \& Tout, C. A. 2000, MNRAS, 317, 773

\item[] Cernicharo, J. \& Reipurth, B. 1996, ApJ, 460, L57

\item[] Eisl\"offel, J., Mundt, R., Ray, T. P., \& Rodr\'\i guez, L. F. 2000, 
in Protostars and Planets IV, ed. V. Mannings, A. P. Boss, \& S. S. Russell 
(Tucson: Univ. Arizona Press), 815

\item[]  Gredel, R. \& Reipurth, B. 1993, ApJ, 407, L29

\item[] Gredel, R. \& Reipurth, B. 1994, A\&A, 289, L19

\item[] Hunter, T.~R., Brogan, 
C.~L., Megeath, S.~T., Menten, K.~M., Beuther, H., 
\& Thorwirth, S.\ 2006, ApJ, 649, 888 

\item[] Lada, C.~J., \& Lada, E.~A.\ 2003, ARA\&A, 41, 57 

\item[] Lefloch, B., Cernicharo, J., Reipurth, B., Pardo, J. R., \& Neri, R.
2007, ApJ, 658, 498

\item[] 
Lim, J., \& Takakuwa, S. 2006, ApJ, 653, 425

\item[] Menten, K. M., Reid, M. J., Forbrich, J., \& Brunthaler, A.
2007, A\&A, 474, 515

\item[] Monin, J.-L., Clarke, C. J., Prato, L., \& McCabe, C. 2007, in 
Protostars and Planets V, ed. B. Reipurth, D. Jewitt, \& K. Keil (Tucson: Univ. Arizona Press),
p. 395

\item[] Nagar, N.M., Vogel, S.N., Stone, J.M., \& Ostriker, E.C. 1997, ApJ 482, L195

\item[] Papaloizou, J. C. B., \& Pringle, J. E. 1987, MNRAS, 225, 267 

\item[] Pichardo, B., Sparke, 
L.~S., \& Aguilar, L.~A.\ 2005, MNRAS, 359, 521 

\item[] Reipurth, B. 1989, Nature, 340, 42

\item[] Reipurth, B., Raga, A.C., \& Heathcote, S. 1992, ApJ 392, 145 

\item[] Reipurth, B., Bally, J., \& Devine, D. 1997, AJ, 114, 2708

\item[] Reipurth, B., Yu, K. C., Rodr\'\i guez, L. F., Heathcote, S., \& Bally, J. 
1999, A\&A, 352, L83

\item[] Rodr\'\i guez, L. F. \& Reipurth, B. 1994, A\&A, 281, 882

\item[] Rodr\'\i guez, L. F., D'Alessio, P., Wilner, D. J., Ho, P. T. P.,
Torrelles, J. M., Curiel, S., G\'omez, Y., Lizano, S., Pedlar, A., Cant\'o, J.,
\& Raga, A. C. 1998, Nature, 395, 355

\item[] Rodr{\'{\i}}guez, L.~F., Porras, A., Claussen, M.~J., Curiel, S., Wilner, 
D.~J., \& Ho, P.~T.~P.\ 2003, ApJ, 586, L137 

\item[] Rodr\'\i guez, L. F., Zapata, L. A., \& Ho, P. T. P.
2007, ApJ, 654, L143

\item[] Rodr\'\i guez, L. F., Moran, J. M., Franco-Hern\'andez, R.,
Garay, G., Brooks, K. J. \& Mardones, D. 2008, AJ, 135, 2370 

\item[] Stapelfeldt, K., Scoville, N.Z. 1993, ApJ, 408, 239

\item[] Yang, J., Ohashi, N., Yan, J., Liu, C., Kaifu, N., \& Kimura, H. 1997, ApJ 475, 683

%\end{thebibliography}
\end{harvard}

\clearpage

%% the file, uncomment the \plotone, \plottwo, and
%% If you need a layout that cannot be achieved with \plotone or
%% \plottwo, you can invoke the graphicx package directly with the
%% \includegraphics command or use \plotfiddle. For more information,
%% please see the tutorial on "Using Electronic Art with AASTeX" in the
%% documentation section at the AASTeX Web site,
%% http://www.journals.uchicago.edu/AAS/AASTeX.
%%
%% The examples below also include sample markup for submission of
%% supplemental electronic materials. As always, be sure to check
%% the instructions to authors for the journal you are submitting to
%% for specific submissions guidelines as they vary from

%% This example uses \plotone to include an EPS file scaled to
%% 80% of its natural size with \epsscale. Its caption
%% has been written to indicate that additional figure parts will be
%% available in the electronic journal.

\begin{figure}
%\epsscale{.70}
\begin{center}
\includegraphics[scale=0.65, angle=0]{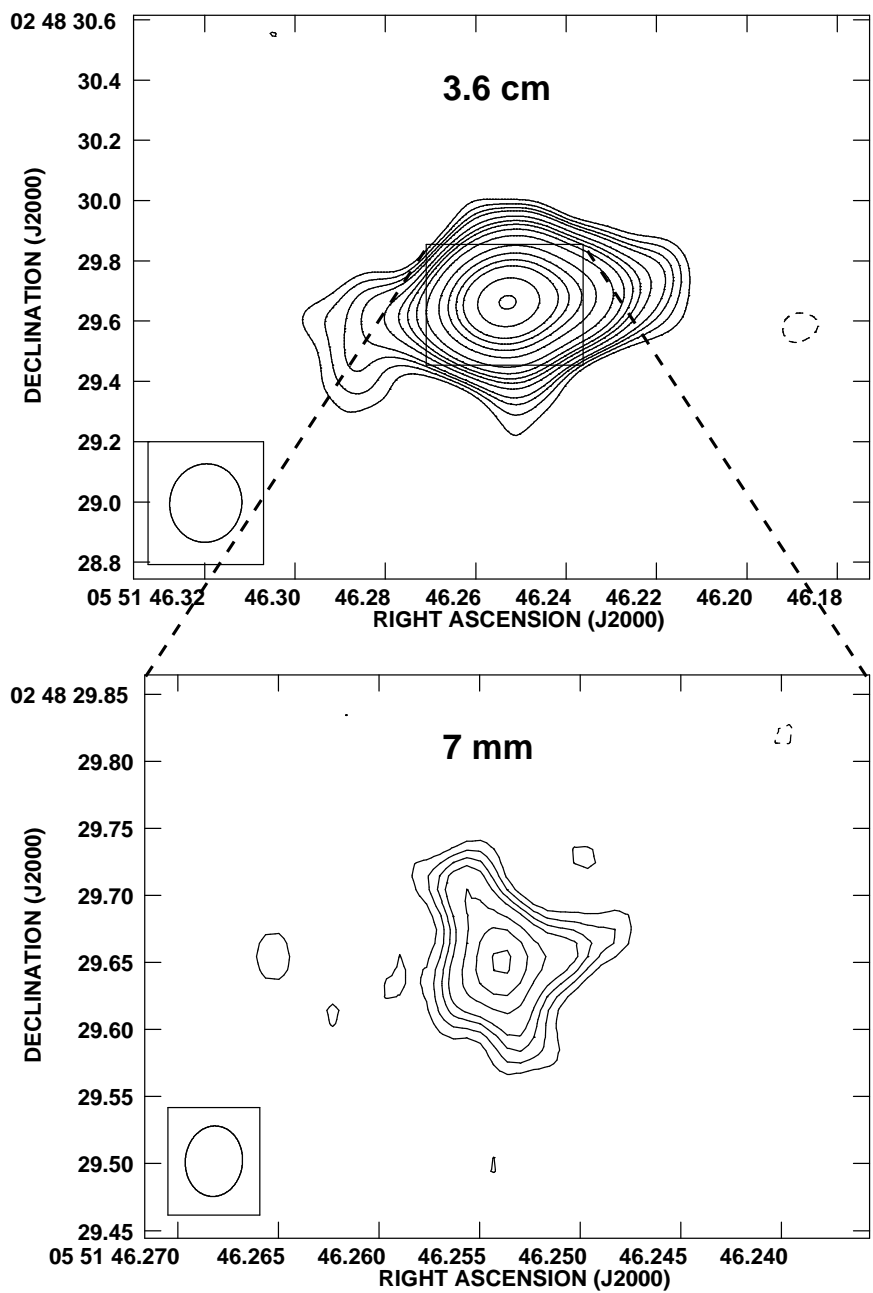}
\end{center}
%\plotone{test2.eps}
\caption{Top: Contour image of the 3.6 cm continuum emission from the core
of the HH111/121 quadrupolar outflow, made from the data of Reipurth et al. (1999).
Contours are
-4, -3, 3, 4, 5, 6, 8, 10, 12, 15, 20 ,30, 40 , 50, 60, 80 and 100
times 4.8 $\mu$Jy beam$^{-1}$, the rms noise of the image. 
The half power contour of the synthesized beam ($0\rlap.{''}26 \times 0\rlap.{''}24$
with a position angle of $-8^\circ$) is shown in the bottom left corner.
The rectangle marks the region shown in the 7 mm image (see below).
Bottom: 
Contour image of the 7 mm continuum emission from the core
of the HH111/121 quadrupolar outflow.
Contours are
-4, -3, 3, 4, 5, 6, 8, 10, and 12
times 56 $\mu$Jy beam$^{-1}$, the rms noise of the image.
The half power contour of the synthesized beam ($0\rlap.{''}053 \times 0\rlap.{''}043$
with a position angle of $-5^\circ$) is shown in the bottom left corner.
Both images were made with the ROBUST parameter of IMAGR set to 0.5.
\label{fig1}}
\end{figure}

\clearpage

\begin{figure}
%\epsscale{.80}
\begin{center}
\includegraphics[scale=0.5, angle=0]{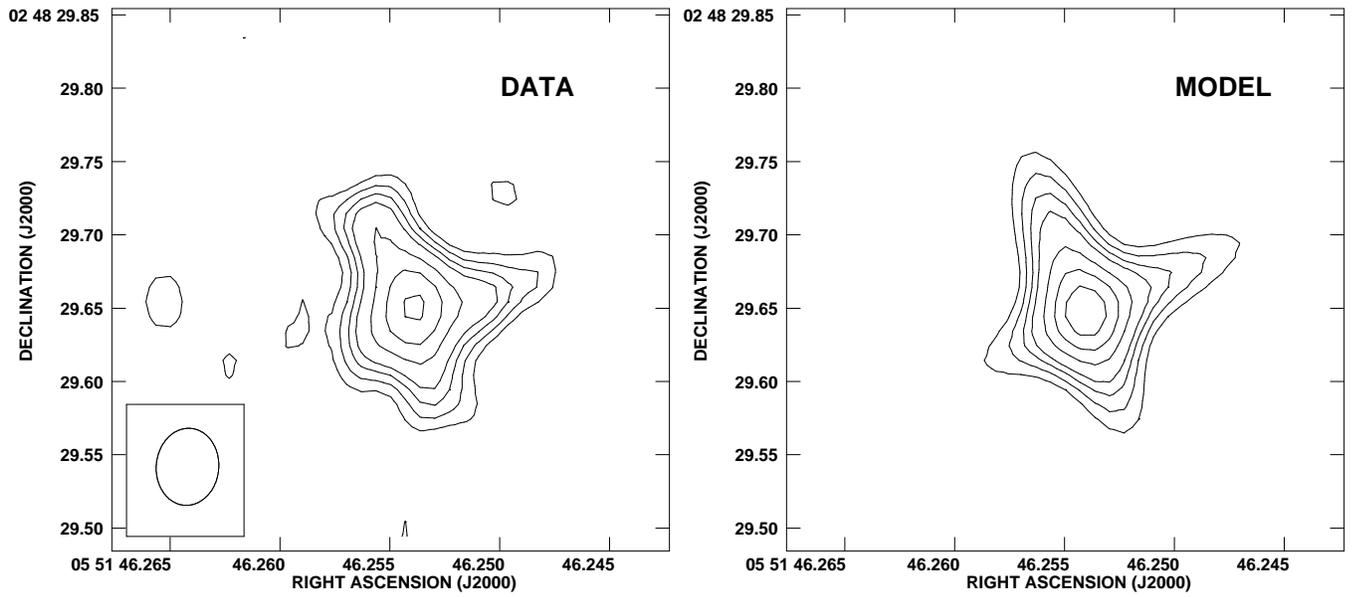}
\end{center}
% \plotone{spectrumnew.eps}
\caption{Contour images of the 7 mm continuum emission (left)
and of the model with two Gaussian ellipsoids.
Contours are
-4, -3, 3, 4, 5, 6, 8, 10, and 12
times 56 $\mu$Jy beam$^{-1}$, the rms noise of the image.
The half power contour of the synthesized beam ($0\rlap.{''}053 \times 0\rlap.{''}043$
with a position angle of $-5^\circ$) is shown in the bottom left corner of the data
image.
\label{fig2}}
\end{figure}

\clearpage

\begin{figure}
%\epsscale{.80}
\begin{center}
\includegraphics[scale=0.75, angle=0]{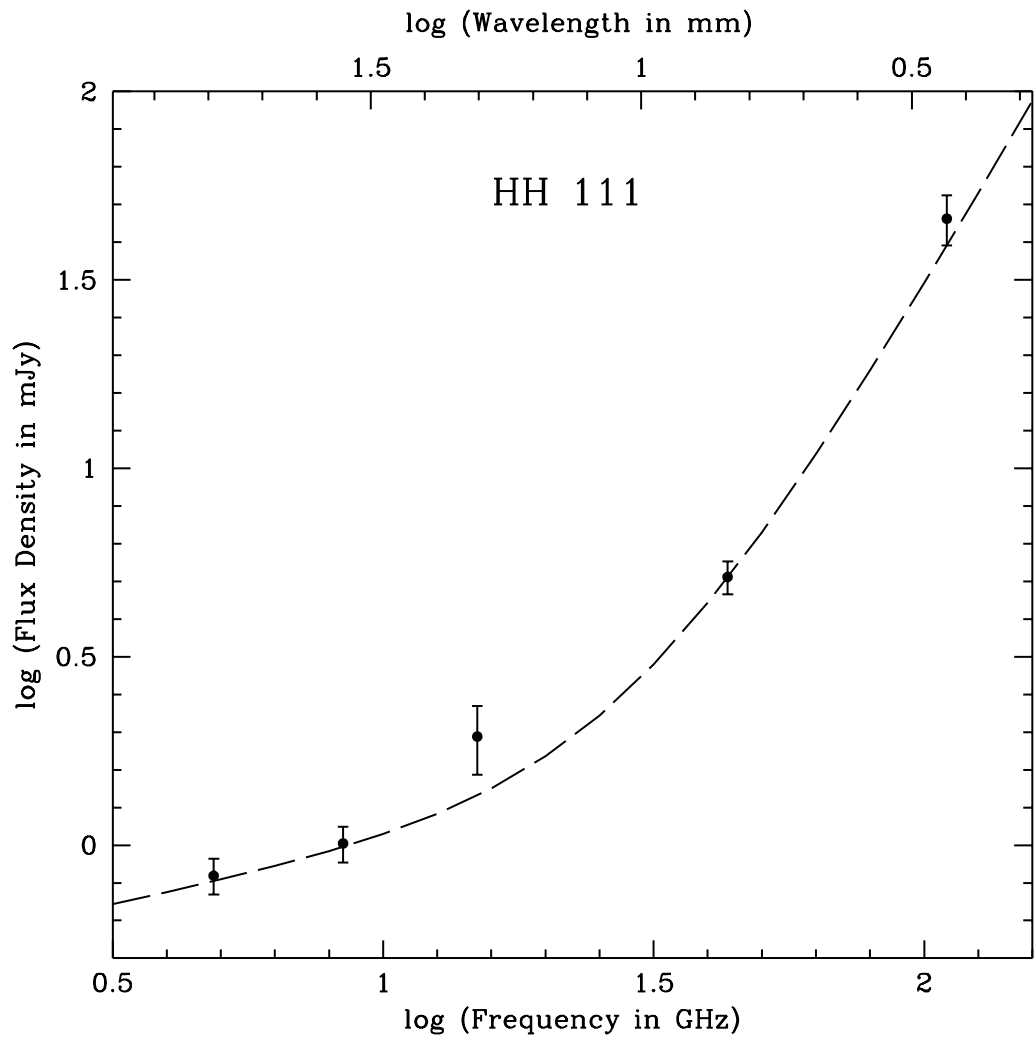}
\end{center}
% \plotone{spectrumnew.eps}
\caption{Continuum data points for the core of the HH 111/121 quadupolar outflow,
from Table 2. The dashed line is the fit described in the text.
\label{fig3}}
\end{figure}

%% Here we use \plottwo to present two versions of the same figure,
%% one in black and white for print the other in RGB color
%% for online presentation. Note that the caption indicates
%% that a color version of the figure will be available online.
%%

%% If you are not including electonic art with your submission, you may
%% mark up your captions using the \figcaption command. See the
%% No more than seven \figcaption commands are allowed per page,
%% so if you have more than seven captions, insert a \clearpage

%% Tables should be submitted one per page, so put a \clearpage before

%% Two options are available to the author for producing tables:  the
%% deluxetable environment provided by the AASTeX package or the LaTeX
%% table environment.  Use of deluxetable is preferred.
%%

%% Three table samples follow, two marked up in the deluxetable environment,
%% one marked up as a LaTeX table.

%% In this first example, note that the \tabletypesize{}
%% command has been used to reduce the font size of the table.
%% We also use the \rotate command to rotate the table to
%% landscape orientation since it is very wide even at the
%% Note also that the \label command needs to be placed

%% This table also includes a table comment indicating that the full
%% version will be available in machine-readable format in the electronic

\clearpage

\begin{table}
\small
%\begin{center}
%\scriptsize
%\small
\caption{Decomposition of the 7 mm Continuum Emission in Two Components.\label{tbl-1}}
%\begin{indented}
%\begin{tabular}{lcccc}
%\tableline\tableline
% &\multicolumn{2}{c}{Position$^a$} & Total Flux
%&  \\
%\cline{2-3}
%Component &  $\alpha$(J2000) & $\delta$(J2000) & Density (mJy) &
%Deconvolved Angular Size$^b$ \\
% \tableline
%\item[]
\begin{tabular}{@{}lcccc}
%\begin{tabular}{lcccc}
\br
 &  & & Total Flux & Deconvolved  \cr
\ns
Component & $\alpha$(J2000)$^a$ & $\delta$(J2000)$^a$ & Density (mJy) & Angular Size$^b$ \cr
\mr
North-South & 05 51 46.2545 & 02 48 29.660 & 2.2$\pm$0.3
& $0\rlap.{''}15 \pm 0\rlap.{''}02 \times 0\rlap.{''}05 \pm 0\rlap.{''}01;~ +23^\circ
\pm 6^\circ$  \\
East-West & 05 51 46.2521 & 02 48 29.659 & 1.4$\pm$0.3
& $0\rlap.{''}20 \pm 0\rlap.{''}05 \times \leq 0\rlap.{''}04;~ +116^\circ
\pm 7^\circ$  \\
\br
\end{tabular}
%% Any table notes must follow the \end{tabular} command.
$^{\rm a}$ Units of right
ascension are hours, minutes, and seconds
and units of declination are degrees, arcminutes, and arcseconds. Absolute positional accuracy
is estimated to be $0\rlap.{''}01$.

$^{\rm b}$ Major axis $\times$ minor axis; position angle of major axis.
%\tablecomments{We can also attach a long-ish paragraph of explanatory
%material to a table.
%\end{indented}
\end{table}
\normalsize

\clearpage

\begin{table}
\small
%\begin{center}
\caption{Total Flux Densities for the Core of the HH 111/121 Quadrupolar Outflow.\label{tbl-2}}
%\begin{tabular}{lcc}
%\begin{indented}
%\item[]\begin{tabular}{@{}lccccc}
\begin{tabular}{lcccccc}
\br
Frequency &  Flux Density & VLA & Angular & VLA & Epoch of & \\
(GHz)  & (mJy) & Configuration & Resolution ($''$) & Project Code & Observation & Reference \\
\br
4.86 & 0.83$\pm$0.09 & CnB & $\sim$4.0 & AA183 & 1994 Sep 21 & VLA archive$^a$  \\
8.44 & 1.01$\pm$0.11 & A & $\sim$0.25 & AR277/8 & 1992 Nov 02 + 1994 Apr 30 & Reipurth et al. (1999)$^a$ \\
14.94 & 1.94$\pm$0.40 & D &  $\sim$5.7 & AR241 & 1991 May 17  & Rodr\'\i guez \& Reipurth (1994)$^a$  \\
43.34 & 5.15$\pm$0.52 & A & $\sim$0.05 & AT325 & 2006 Feb 10+15 & This paper  \\
110.2 & 46.0$\pm$7.0 & --$^b$ & $\sim$7.0 & --$^b$ & 1989 Dec - 1990 May & Stapelfeldt \& Scoville (1993)  \\
\br
\end{tabular}
%% Any table notes must follow the \end{tabular} command.
$^{\rm a}$ Flux densities from our analysis of the data.

$^{\rm b}$ Data taken with the Owens Valley Radio Observatory Millimeter Interferometer.
%\tablenotetext{a}{Units of right 
%ascension are hours, minutes, and seconds
%and units of declination are degrees, arcminutes, and arcseconds. Positional accuracy
%is estimated to be $0\rlap.{''}05$.}
%\tablecomments{We can also attach a long-ish paragraph of explanatory
%material to a table.}
%\end{indented}
\end{table}
\normalsize

%If the table is more than one page long, the width of the table can vary
%% from page to page when the default \tablewidth is used, as below.  The
%% individual table widths for each page will be written to the log file; a
%% maximum tablewidth for the table can be computed from these values.
%% The \tablewidth argument can then be reset and the file reprocessed, so
%% that the table is of uniform width throughout. Try getting the widths
%% from the log file and changing the \tablewidth parameter to see how
%% adjusting this value affects table formatting.

%% The \dataset{} macro has also been applied to a few of the objects to
%% show how many observations can be tagged in a table.

%% Tables may also be prepared as separate files. See the accompanying
%% sample file table.tex for an example of an external table file.
%% To include an external file in your main document, use the \input
%% command. Uncomment the line below to include table.tex in this
%% sample file. (Note that you will need to comment out the \documentclass,
%% \begin{document}, and \end{document} commands from table.tex if you want
%% to include it in this document.)


%% The following command ends your manuscript. LaTeX will ignore any text

\end{document}